\documentclass[10pt,conference]{IEEEtran}
\IEEEoverridecommandlockouts
\usepackage{cite}
\usepackage{amsmath,amssymb,amsfonts}
\usepackage{algorithmic}
\usepackage{graphicx}
\usepackage{textcomp}
\usepackage{xcolor}
\usepackage{tabularx}
\usepackage{xurl}
\usepackage[hidelinks, breaklinks=true]{hyperref}
\usepackage{multirow}
\usepackage{booktabs}
\usepackage{comment}
\usepackage{caption}
\usepackage{tabularray}
\usepackage{dblfloatfix}  
\usepackage[section]{placeins}  
\def\BibTeX{{\rm B\kern-.05em{\sc i\kern-.025em b}\kern-.08em
    T\kern-.1667em\lower.7ex\hbox{E}\kern-.125emX}}
\begin{document}



\title{AFSS: Artifact-Focused Self-Synthesis for Mitigating Bias in Audio Deepfake Detection}


\author{
\begin{tabular}{c}
Hai-Son Nguyen-Le$^{\star \dagger}$ \qquad
Hung-Cuong Nguyen-Thanh$^{\star \dagger}$ \qquad
Nhien-An Le-Khac$^{\ddagger}$ \\[0.3em]
Dinh-Thuc Nguyen$^{\dagger}$ \qquad
Hong-Hanh Nguyen-Le$^{\ddagger}$
\end{tabular}
\\[1em]
\small
$^{\star}$ Two authors have the equal contribution \\
$^{\dagger}$ University of Science, Ho Chi Minh City, Vietnam \\
$^{\ddagger}$ University College Dublin, School of Computer Science, Dublin, Ireland
}

\maketitle

\begin{abstract}
The rapid advancement of generative models has enabled highly realistic audio deepfakes, yet current detectors suffer from a critical bias problem, leading to poor generalization across unseen datasets. This paper proposes Artifact-Focused Self-Synthesis (AFSS), a method designed to mitigate this bias by generating pseudo-fake samples from real audio via two mechanisms: self-conversion and self-reconstruction. The core insight of AFSS lies in enforcing same-speaker constraints, ensuring that real and pseudo-fake samples share identical speaker identity and semantic content. This forces the detector to focus exclusively on generation artifacts rather than irrelevant confounding factors. Furthermore, we introduce a learnable reweighting loss to dynamically emphasize synthetic samples during training. Extensive experiments across 7 datasets demonstrate that AFSS achieves state-of-the-art performance with an average EER of 5.45\%, including a significant reduction to 1.23\% on WaveFake and 2.70\% on In-the-Wild, all while eliminating the dependency on pre-collected fake datasets. Our code is publicly available at \url{https://github.com/NguyenLeHaiSonGit/AFSS}.
\end{abstract}


\section{Introduction}
The rapid development of generative models (GMs), particularly Generative Adversarial Networks (GANs) \cite{kong2020hifi}, Variational Autoencoders (VAEs) \cite{qian2020unsupervised}, or Diffusion Models (DMs) \cite{huang2023make}, has enabled the creation of highly realistic synthetic speech. Voice conversion (VC) and text-to-speech (TTS) systems built upon these architectures can now generate audio that is perceptually indistinguishable from genuine human speech \cite{kumar2023deep}. While these technologies offer legitimate applications in entertainment and human-computer interaction, they have been increasingly exploited for malicious purposes, such as financial fraud, identity impersonation, and large-scale misinformation campaigns \cite{suwajanakorn2017synthesizing, stupp2019, surasit2024}. The sophistication and accessibility of these synthesis tools underscore the urgent need for effective audio deepfake (DF) detection methods capable of identifying manipulated audio across diverse real-world scenarios.

\begin{figure}[ht]
  \centering
  \includegraphics[width=0.9\columnwidth]{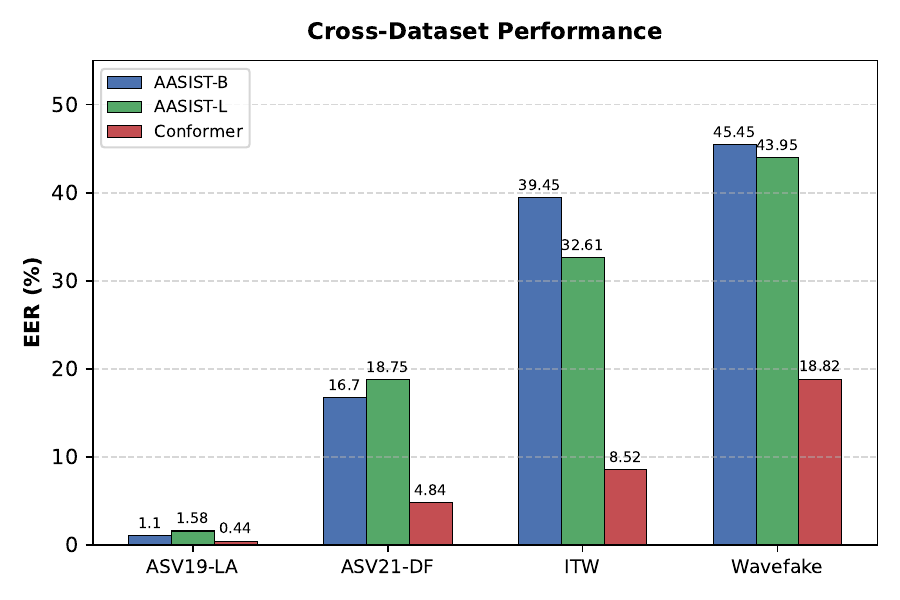}
  \captionsetup{}
  \caption{Cross-dataset performance of audio deepfake detection methods (EER\%) when trained on ASV19-LA and tested on different datasets. Lower values indicate better performance.}
  \label{fig:poor-generalization}
\end{figure}

Recent detection approaches can be broadly categorized into two architectural paradigms. The first employs specialized end-to-end (E2E) architectures that learn discriminative features directly from spectral representations. Notable examples include RawNet \cite{rawnet2_14, rawnet2_15}, which performs time-domain convolution directly on raw audio, and AASIST \cite{jung2022aasist}, which utilizes graph neural networks to model the non-Euclidean relationships within speech signals to capture complex short-range and long-range features. These methods design task-specific architectures optimized for capturing synthesis artifacts. The second paradigm adopts self-supervised learning (SSL) models as front-end feature extractors. Models such as Wav2vec 2.0 \cite{schneider2019wav2vec}, XLS-R \cite{babu2021xls}, and HuBERT \cite{hsu2021hubert} are first pre-trained on large-scale generic speech corpora to learn rich acoustic representations, then fine-tuned on downstream DF detection tasks. This paradigm has gained popularity due to its ability to leverage powerful pre-trained representations that capture subtle acoustic characteristics directly from raw waveforms, eliminating the need for handcrafted features such as MFCCs or spectrograms.


Despite achieving excellent in-domain accuracy, both paradigms exhibit significant performance degradation when evaluated on unseen datasets, as shown in Fig. \ref{fig:poor-generalization}. This considerable cross-domain performance gap reveals a fundamental limitation of current detection approaches that transcends architectural choices. The primary cause of this failure is the \emph{\textbf{bias problem}}: detectors inadvertently learn to distinguish real from fake audio based on irrelevant information rather than authentic forgery artifacts \cite{yan2023ucf}. For SSL-based approaches, models like Wav2vec 2.0 inherently encode semantic content and speaker identity since they were optimized for speech recognition and speaker identification tasks. For E2E architectures, the bias emerges from dataset-specific characteristics such as speaker populations or recording conditions. When these irrelevant features vary across datasets, detection performance degrades substantially \cite{lu2024one}. In addition to the bias problem, collecting large and diverse fake datasets for training remains a major practical challenge. Although recent studies, such as the ALDEN framework \cite{xu2025alden}, have attempted to disentangle synthetic-irrelevant information to improve generalization, these methods often still require extensive fake data from the training dataset. This dependency makes it difficult for detection systems to keep pace with the continuous emergence of new generative technologies.

In this work, we propose \textbf{\emph{Artifact-Focused Self-Synthesis (AFSS)}} method specifically designed to mitigate the bias problem in audio DF detection. The core insight of AFSS is to generate pseudo-fake samples that contain authentic forgery artifacts while explicitly controlling for confounding factors. AFSS employs two artifact-focused self-generation mechanisms: \textbf{self-conversion}, which transforms audio using the same speaker as both source and target, and \textbf{self-reconstruction}, which passes audio through neural vocoders to introduce characteristic synthesis artifacts. By ensuring that real and pseudo-fake samples share identical speaker identity and semantic content, AFSS eliminates the possibility of the detector exploiting irrelevant features (i.e., speakers' identity) as classification shortcuts. The detector is thus forced to focus exclusively on the generation artifacts introduced by VC systems and neural vocoders. Furthermore, we introduce a learnable reweighting loss that dynamically adjusts the emphasis on synthetic samples during training, encouraging the model to actively explore and learn universal generation artifacts for enhanced cross-domain generalization. Importantly, AFSS operates without requiring any pre-collected fake datasets, generating all training samples from real audio alone. 

Our main contributions are summarized as follows:
\begin{itemize}
    \item We propose Artifact-Focused Self-Synthesis method to mitigate the bias problem in audio DF detection by generating artifact-focused pseudo-fake samples through self-conversion and self-reconstruction mechanisms with same-speaker constraints. Our method also eliminates dependency on pre-collected fake datasets by generating all training samples from real audio alone.
    \item We introduce a learnable reweighting loss that dynamically emphasizes synthetic samples, encouraging the model to focus on universal generation artifacts.
    \item Extensive experiments across seven datasets demonstrate state-of-the-art cross-domain performance with an average EER $5.45\%$, including $1.23\%$ on WaveFake and $2.70\%$ on In-the-Wild.
\end{itemize}


\section{Related Work}
\subsection{Audio Deepfake Detection Methods}
Recent advancements in audio DF detection are primarily characterized by two architectural paradigms: the pipeline-based and the E2E detection approaches. 

Pipeline-based approaches employ a two-stage architecture where the front-end feature extractor operates independently of the back-end classifier. The front-end computes handcrafted acoustic features such as MFCCs, LFCCs, or spectrograms through signal processing algorithms without learnable parameters. These fixed representations are then fed to a separately trained back-end classifier. Since no gradient flows from the classifier to the feature extractor, the two stages are optimized in isolation. In contrast, E2E approaches integrate feature extraction and classification into a unified network where backpropagation flows from the classification output through the entire architecture \cite{rawnet2_14, rawnet2_15, jung2022aasist, doan2024balance, rosello2023conformer}. This enables joint optimization of both components for the detection objective. Furthermore, SSL-based methods \cite{tak2022automatic, wang2024can, conneau2020unsupervised} using Wav2vec 2.0 or XLS-R as fine-tuned front-ends also fall within this paradigm, as gradient updates propagate through the pre-trained encoder during task-specific training.

Beyond architectural design, several strategies address generalization: one-class learning models bonafide distributions to detect anomalies \cite{kim2024one}; synthetic data training exposes detectors to diverse TTS and VC outputs \cite{wang2023spoofed}, \cite{wang2024can}; augmentation techniques such as RawBoost \cite{tak2022rawboost} simulate acoustic distortions; and parameter-efficient methods including LoRA enable adaptation to novel generators \cite{wu2024adapter}.

\subsection{Disentanglement methods for audio DF detection}

Disentanglement methods for audio DF detection seeks to separate forgery-relevant from forgery-irrelevant factors to mitigate bias in audio deepfake detection. DSVAE \cite{yadav2023dsvae} introduces an interpretable framework using a Variational Autoencoder to decompose speech into recording environments and fundamental speech characteristics, enhancing robustness against unseen synthesizers. Similarly, SafeEar \cite{li2024safeear} proposes a content-privacy-preserving mechanism that disentangles audio into "forgery-related" acoustic features and "identity-related" semantic features, utilizing adversarial training to focus the detector solely on acoustic artifacts. More recently, the ALDEN framework \cite{xu2025alden} has emerged as one of the leading methods in the field by expanding this concept into a dual-level architecture. By isolating vocoder-specific signals from high-level semantics, ALDEN achieves high generalization performance in real-world scenarios.

While effective, these approaches share two fundamental limitations: (1) they require pre-collected fake datasets to learn the disentanglement, creating dependency on existing synthetic corpora, and (2) they rely on the network's capacity to implicitly separate confounding factors during training, which may be dataset-specific. AFSS takes a fundamentally different approach by eliminating confounding factors by construction rather than learning to disentangle them. Through same-speaker constraints, real and pseudo-fake samples share identical speaker identity and semantic content by design, guaranteeing that these factors cannot serve as classification shortcuts. This explicit constraint removes dependency on pre-collected fake datasets while ensuring complete bias elimination.

\section{PROPOSED METHOD}
This section details our proposed \textbf{Artifact-Focused Self-Synthesis (AFSS)} method. The core insight of AFSS is that the bias problem can be systematically eliminated by ensuring that real and pseudo-fake samples share identical speaker identity and semantic content. Under this constraint, the detector cannot exploit speaker-related or content-related features as classification shortcuts, and is thus forced to focus exclusively on the generation artifacts introduced by VC systems and neural vocoders. 

\begin{figure}[ht]
  \centering
  \includegraphics[width=0.99\columnwidth]{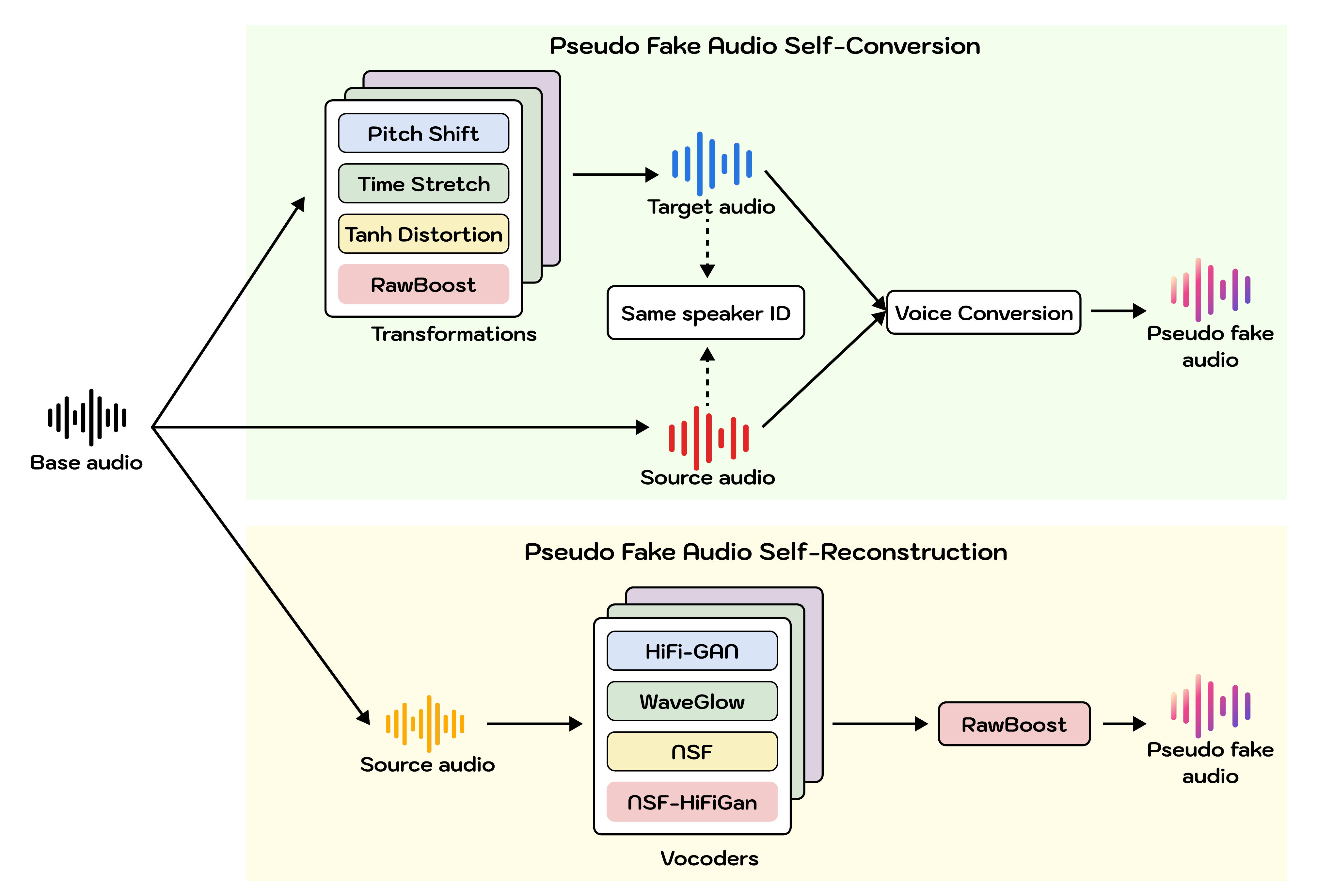}
  \captionsetup{}
  \caption{Overview of the proposed AFSS method.}
  \label{fig:pseudofakegen}
\end{figure}

\begin{figure*}[ht]
  \centering
  \includegraphics[width=\textwidth]{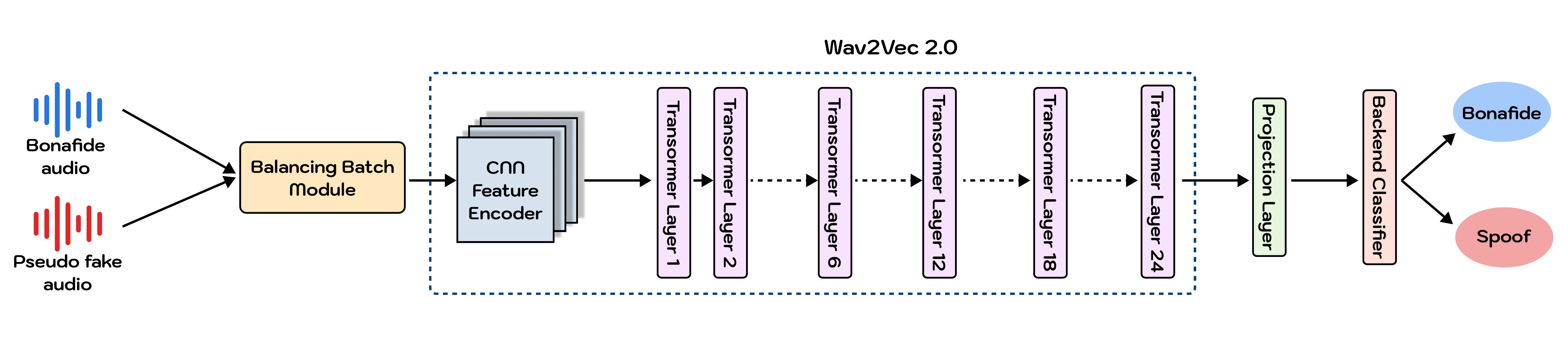}
  \captionsetup{}
  \caption{Overview of model architecture.}
  \label{fig:modelarchitecture}
\end{figure*}

AFSS operates through two complementary branches: self-conversion and self-reconstruction, with each branch targets distinct categories of forgery artifacts. Fig. \ref{fig:pseudofakegen} illustrates our proposed method, while Fig. \ref{fig:modelarchitecture} presents the overview of our model architecture. 

\subsection{Pseudo-Fake Audio Self-Conversion}

In conventional training paradigms, VC systems transform audio from a source speaker to a target speaker, creating a systematic correlation between speaker identity changes and the fake label \cite{yi2023audio}. Detectors trained on such data may learn to identify \textit{speaker inconsistency} rather than actual VC artifacts, leading to poor generalization when encountering new speaker populations. Our self-conversion branch addresses this bias by converting from the same speaker identity. 

Given a base audio sample $x_b$ from speaker $s$, we first generate an acoustically modified variant $x_t = \alpha_i(x_b)$ using transformation techniques $\mathcal{A} = \{\alpha_i\}_{i=1}^{m}$, where each $\alpha_i$ represents a specific acoustic modification (i.e., pitch shifting, time stretching, tanh distortion, or RawBoost \cite{tak2022rawboost}. Crucially, $\alpha_i$ preserves speaker identity while altering acoustic characteristics sufficient to drive the VC process. The VC module $\mathcal{V}$ then transforms the source audio to match the target: $\hat{x}_{\text{vc}} = \mathcal{V}(x_b, x_t)$.
Since both $x_b$ and $x_t$ originate from the same speaker $s$, the resulting pseudo-fake audio $\hat{x}_{\text{vc}}$ contains authentic VC artifacts  without any speaker identity confounders. 

This approach is compatible with any VC system. In our implementation, we employ kNN-VC \cite{baas2023voice} for its computational efficiency and lightweight architecture, though the framework generalizes to arbitrary VC models.

\textbf{Transformation Strategy.} The placement of transformation techniques before the VC process is critical for artifact preservation. For each sample, we apply one of the four transformation techniques, including PitchShift, TimeStretch, TanhDistortion, or RawBoost, which are selected randomly with equal probability. These applied techniques ensure sufficient acoustic diversity to drive the conversion module while avoiding excessive signal distortion that could compromise speaker identity.

\subsection{Pseudo-Fake Audio Self-Reconstruction}
The self-reconstruction branch targets a complementary source of bias: the characteristic fingerprints left by neural vocoders used in the final stage of most TTS pipelines. To expose the detector to diverse vocoder artifacts while maintaining speaker and content consistency, we process base audio through various vocoders $\mathcal{G} = \{g_i\}_{i=1}^{n}$ for reconstruction: $\hat{x}_{\text{rec}} = g_i(\text{encode}(x_b))$.

This simulates the encoding-decoding pipeline typical of DF generation, introducing characteristic vocoder artifacts including over-smoothed spectral details, phase inconsistencies, and loss of natural micro-variations. Since the reconstructed audio $\hat{x}_{\text{rec}}$ derives from the original $x_b$, speaker identity and semantic content remain unchanged. The framework supports any reconstruction
vocoder, but in our experiment, we select four widely-used non-autoregressive models: HiFiGAN \cite{kong2020hifi}, WaveGlow \cite{prenger2019waveglow}, Hn-NSF \cite{wang2019neural}, and NSF-HiFiGAN \cite{tomashenko2024voiceprivacy}.

\textbf{Transformation Strategy.} Unlike the self-conversion branch, transformations in this branch are applied after the vocoder reconstruction using exclusively RawBoost. This placement is physically motivated: neural vocoder fingerprints reside in delicate micro-variations of spectral and phase detail. Aggressive transformations such as PitchShift or TimeStretch applied before or during reconstruction would mask these subtle traces. This design ensures the detector learns robust vocoder fingerprints that generalize across transmission conditions.


\subsection{Learnable Reweighting Loss Function}
The standard Binary Cross-Entropy loss treats all authentic and synthetic audio with equal importance, which prevents the model from adequately focusing on synthetic samples generated by our AFSS mechanism. In this work, we expect the model to explore artifacts left by VC models and vocoders by emphasizing synthetic samples, helping enhance cross-domain generalization. To achieve this, we propose a Learnable Reweighting Loss:
\begin{equation}
    \mathcal{L} = -[w_{fake} \cdot y \log(p) + w_{real} \cdot (1-y) \log(1-p)],
\end{equation}
where $y \in \{0,1\}$ is the ground-truths and $p$ is the predicted priori probability. Unlike fixed-weight approaches, $w_{fake}$ and $w_{real}$ are derived from learnable parameters $\tilde{w}_{real}$ and $\tilde{w}_{fake}$ optimized via backpropagation. Sigmoid transformations ensure stable weight ranges:
\begin{equation}
    w_{\text{fake}} = 1.0 + \sigma(\tilde{w}_{fake}), \quad w_{\text{real}} = \sigma(\tilde{w}_{real})
\end{equation}
where $\sigma$ denotes the sigmoid function. This formulation strictly constrains $w_{\text{fake}} \in (1,2)$ and $w_{\text{real}} \in (0,1)$. The \textbf{1.0 bias} for $w_fake$ ensures that self-generated samples always receive higher emphasis than real samples. This setup acts as a dynamic hard-mining mechanism: when the model encounters "hard" synthetic samples, the learnable weights automatically adjust to prioritize their gradients.


\subsection{Balanced Batch Sampling}

Random batch sampling can introduce class imbalance that destabilizes training and interferes with our learnable loss function. Inspired by the work \cite{doan2024balance}, we implement a balanced batch strategy that ensures each mini-batch contains equal numbers of real and fake samples. This strategy prevents the learnable weights from being skewed by random class distribution fluctuations, maintaining stable gradient updates and consistent learning dynamics throughout training.

\section{EXPERIMENTAL SETUP}

\subsection{Dataset}

\subsubsection{Training Dataset}
Since our AFSS method automatically generates pseudo-fake samples from real audio, we only use authentic audio samples from the training and development partitions of the ASVspoof 2019 LA dataset \cite{wang2020asvspoof} for training.

\subsubsection{Evaluation Datasets}

To evaluate our method on cross-domain generation scenarios, we utilize 7 different datasets: ASVspoof 2019 LA-Eval \cite{wang2020asvspoof}, ASVspoof 2021 LA evaluation and hidden sets \cite{liu2023asvspoof}, ASVspoof-2021 DF evaluation and hidden sets \cite{liu2023asvspoof}, WaveFake \cite{frank2021wavefake}, and In-the-Wild \cite{muller2022does}.



\subsection{Baseline}

We compare our method with 5 baselines, including AASIST-B \cite{jung2022aasist}, AASIST-L \cite{jung2022aasist}, SCL \cite{doan2024balance}, Conformer-based detector \cite{rosello2023conformer}, and SSL \cite{wang2024can}. We use the code provided by these studies for implementation, excluding the work \cite{wang2024can} which we report the results from the authors' publication. We also compare with disentanglement methods: ALDEN \cite{xu2025alden}, DSVAE \cite{yadav2023dsvae}, and SafeEar \cite{li2024safeear}. 

\subsection{Evaluation Metric}
We use Equal Error Rate (EER), Accuracy (ACC), Area Under the ROC Curve (AUC), and Average Precision (AP) as the primary evaluation metrics for this paper.

\subsection{Implementation Details}
The framework utilizes XLS-R as the front-end feature extractor, projecting 1024-dimensional features to 128 dimensions via a linear layer. The back-end classifier comprises ReLU activation, Dropout (p = 0.5), Mean Pooling, and a Dense layer for final classification.

Training employs AdamW optimizer with weight decay of $10^{-4}$ for 30 epochs maximum. We use differential learning rates: $5 \times 10^{-6}$ for the SSL front-end, $10^{-4}$ for back-end layers and learnable loss parameters. A 5-epoch linear warm-up precedes linear decay to $10^{-6}$. BalancedBatchSampler creates mini-batches of 12 samples (6 bona fide, 6 fake). Early stopping with patience of 10 epochs ensures training stability.

\subsection{Transformation Intensities}
The intensity levels are defined by the aggressiveness of their parameter ranges:
\begin{itemize}
    \item Intensity 0 (Baseline): Applies minimal transformation.
    \item Intensity 1 (Optimal): Our proposed setting that balances diversity and realism. It includes TimeStretch ($[0.9, 1.1]$), TanhDistortion ($[0.15, 0.6]$), PitchShift ($[-0.5, 0.5]$ semitones), and RawBoost (utilizing both convolutive noise with $nBands=5$ and additive noise).
    \item Intensity 2 \& 3: Use increasingly aggressive ranges to simulate highly distorted environments.
\end{itemize}

\section{Experimental Results}

\subsection{Performance Across Datasets}
We report results on EER and AP metrics across 7 audio DF datasets. From Tab. 
\ref{tab:main_eer} and \ref{tab:main_auc}, some insights can be observed:


\begin{table*}[!ht]
    \centering
    \small
    \setlength{\tabcolsep}{5pt}
    \begin{tabular}{l l c cccc cc |c}
    \toprule
    & & \textbf{ASVspoof 19} & \multicolumn{4}{c}{\textbf{ASVspoof 21}} & \multicolumn{2}{c}{\textbf{Real-World}} & \\
    \cmidrule(lr){3-3} \cmidrule(lr){4-7} \cmidrule(lr){8-9}
    \textbf{Method} & \textbf{Venue} & eval & DF eval & DF hidden & LA eval & LA hidden & In-the-Wild & WaveFake & \textbf{Avg} \\
    \midrule
    AASIST-B \cite{jung2022aasist} & ICASSP '22 & 1.10 & 16.70 & 21.16 & 7.40 & 32.81 & 39.45 & 45.45 & 23.44 \\
    AASIST-L \cite{jung2022aasist} & ICASSP '22 & 1.58 & 18.75 & 20.11 & 8.57 & 26.47 & 32.61 & 43.95 & 21.72 \\
    SCL \cite{doan2024balance} & Interspeech '24 & 2.88 & \textbf{2.17} & \textbf{5.76} & 13.90 & 15.15 & 4.52 & 2.99 & 6.77 \\
    Conformer \cite{rosello2023conformer} & Interspeech '23 & \textbf{0.44} & 4.84 & 8.37 & \textbf{1.67} & \textbf{12.19} & 8.52 & 18.82 & 7.84 \\
    SSL \cite{wang2024can} & ICASSP '24 & 1.91 & 5.67 & 8.84 & 15.92 & 14.97 & 6.10 & 1.30 & 7.82 \\
    \midrule
    \textbf{AFSS (Ours)} &  & 2.72 & 2.19 & 6.92 & 10.02 & 12.35 & \textbf{2.70} & \textbf{1.23} & \textbf{5.45} \\
    \bottomrule
    \end{tabular}
    \caption{Cross-dataset generalization performance comparison (EER\%). Lower is better. Best results are in \textbf{bold}.}
    \label{tab:main_eer}
\end{table*}

\begin{table*}[!ht]
    \centering
    \small
    \setlength{\tabcolsep}{5pt}
    \begin{tabular}{l l c cccc cc |c}
    \toprule
    & & \textbf{ASVspoof 19} & \multicolumn{4}{c}{\textbf{ASVspoof 21}} & \multicolumn{2}{c}{\textbf{Real-World}} & \\
    \cmidrule(lr){3-3} \cmidrule(lr){4-7} \cmidrule(lr){8-9}
    \textbf{Method} & \textbf{Venue} & eval & DF eval & DF hidden & LA eval & LA hidden & In-the-Wild & WaveFake & \textbf{Avg} \\
    \midrule
    AASIST-B \cite{jung2022aasist} & ICASSP '22 & 0.9984 & 0.9059 & 0.8389 & 0.9811 & 0.7229 & 0.6531 & 0.9165 & 0.8595 \\
    AASIST-L \cite{jung2022aasist} & ICASSP '22 & 0.9982 & 0.8989 & 0.8652 & 0.9702 & 0.7985 & 0.7372 & 0.9250 & 0.8847 \\
    SCL \cite{doan2024balance} & Interspeech '24 & 0.9984 & \textbf{0.9968} & \textbf{0.9851} & 0.9343 & 0.9200 & 0.9910 & 0.9901 & 0.9737 \\
    Conformer \cite{rosello2023conformer} & Interspeech '23 & \textbf{0.9995} & 0.9903 & 0.9720 & \textbf{0.9942} & 0.9418 & 0.9712 & 0.8868 & 0.9651 \\
    \midrule
    \textbf{AFSS (Ours)} & & 0.9939 & 0.9965 & 0.9794 & 0.9610 & \textbf{0.9440} & \textbf{0.9964} & \textbf{0.9990} & \textbf{0.9815} \\
    \bottomrule
    \end{tabular}
    \caption{Cross-dataset generalization performance comparison (AUC). Higher is better. Best results are in \textbf{bold}.}
    \label{tab:main_auc}
\end{table*}

\begin{table*}[ht]
    \centering
    \small
    \setlength{\tabcolsep}{5pt}
    \begin{tabular}{l l c cccc cc}
    \toprule
    & & \textbf{ASVspoof 19} & \multicolumn{4}{c}{\textbf{ASVspoof 21}} & \multicolumn{2}{c}{\textbf{Real-World}} \\
    \cmidrule(lr){3-3} \cmidrule(lr){4-7} \cmidrule(lr){8-9}
    \textbf{Method} & \textbf{Venue} & LA eval & DF eval & DF hidden & LA eval & LA hidden & In-the-Wild & WaveFake \\
    \midrule
    ALDEN \cite{xu2025alden} & ACM MM '25 & 1.22 & 4.39 & - & - & - & 11.27 & 4.66 \\
    DSVAE \cite{yadav2023dsvae} & & \textbf{2.16} & - & - & - & - & - & - \\
    SafeEar \cite{li2024safeear} & CCS '24 & 3.10 & - & - & \textbf{7.22} & - & - & - \\
    \midrule
    \textbf{AFSS (Ours)} & & 2.72 & \textbf{2.19} & 6.92 & 10.02 & 12.35 & \textbf{2.70} & \textbf{1.23} \\
    \bottomrule
    \end{tabular}
    \caption{Comparison with disentanglement methods (EER\%). Results are taken from original publications; '-' indicates the metric was not reported by the original authors. Best results are in \textbf{bold}.}
    \label{tab:disentanglement_eer}
\end{table*}


\begin{itemize}
    \item \textbf{Superior Cross-Domain Generalization}. AFSS achieves the lowest average EER of $5.45\%$ and highest average AUC of $98.15\%$ across all evaluation datasets. This validates our core hypothesis: eliminating speaker identity and semantic content as confounding factors forces the detector to learn universal generation artifacts.
    \item \textbf{High Real-World Performance}. The most pronounced advantages appear on challenging real-world scenarios. For example, on In-the-Wild dataset, AFSS achieves 2.70\% EER, substantially outperforming AASIST-L (32.61\%) and Conformer (8.52\%). These results demonstrate effective transfer to unseen synthesis methods.
    \item \textbf{Generalization-Specialization Trade-off}. While Conformer achieves lower EER on ASVspoof 2019-eval, this advantage does not transfer to out-of-domain scenarios. AFSS sacrifices marginal in-domain accuracy for substantially better generalization, which is a more desirable property for real-world deployment where attack methods are unknown a priori.
\end{itemize}

\subsection{Comparison with Disentanglement Methods}

Tab.\ref{tab:disentanglement_eer} compares AFSS against state-of-the-art disentanglement approaches. While ALDEN achieves lower EER on ASVspoof 2019-eval (1.22\% vs. 2.72\%), AFSS demonstrates superior performance on challenging real-world datasets: $4.2\times$ improvement on In-the-Wild and $3.8\times$ improvement on WaveFake. The key distinction lies in methodology. Disentanglement methods such as ALDEN, DSVAE, and SafeEar require pre-collected fake datasets to learn representations that separate forgery-relevant from forgery-irrelevant features. In contrast, AFSS eliminates this dependency entirely by generating pseudo-fake samples from real audio alone. 

\subsection{Ablation Study and Analysis}
\subsubsection{Components Contribution}
Tab.\ref{tab:component_effectiveness} shows that combining self-conversion and self-reconstruction (S-VC + S-Rec) achieves 8.57\% EER on ASVspoof 2019-eval and 4.08\% on In-the-Wild. This result validates that same-speaker constraints alone provide effective bias mitigation. Combining the learnable reweighting loss with balanced batch sampling yields improvements, reducing EER from 8.57\% to 2.72\% on ASVspoof 2019-eval and from 4.08\% to 2.70\% on In-the-Wild. 

\begin{table}[!t]
    \centering
    \footnotesize
    \setlength{\tabcolsep}{2.5pt}
    \begin{tabular}{lcccccccc}
    \toprule
    \textbf{Components} & \multicolumn{4}{c}{\textbf{ASV19-eval}} & \multicolumn{4}{c}{\textbf{In-the-Wild}} \\
    \cmidrule(lr){2-5} \cmidrule(lr){6-9}
     & \textbf{EER} & \textbf{AUC} & \textbf{ACC} & \textbf{AP} & \textbf{EER} & \textbf{AUC} & \textbf{ACC} & \textbf{AP} \\
    \midrule
    S-VC + S-Rec & 8.57 & 97.81 & 91.44 & 99.75 & 4.08 & 99.14 & 95.92 & 98.57 \\
    + Alpha & 6.61 & 97.37 & 93.39 & 99.71 & 3.19 & 99.52 & 96.81 & 99.19 \\
    + Bal & 6.84 & 97.31 & 93.16 & 99.69 & 3.74 & 99.29 & 96.26 & 98.81 \\
    \textbf{+ Alpha + Bal} & \textbf{2.72} & \textbf{99.39} & \textbf{97.26} & \textbf{99.93} & \textbf{2.70} & \textbf{99.64} & \textbf{97.30} & \textbf{99.39} \\
    \bottomrule
    \end{tabular}
    \caption{Effectiveness of the proposed components (\%). S-VC: Self-conversion, S-Rec: Self-reconstruction, Alpha: Learnable Reweighting Loss, Bal: Balanced Batch Sampling.}
    \label{tab:component_effectiveness}
\end{table}



\subsubsection{Effects of Transformation Intensity Levels}

\begin{table}[ht]
    \centering
    \resizebox{\columnwidth}{!}{%
    \begin{tabular}{lcccc}
    \toprule
    \textbf{Dataset} & \textbf{Intensity 0} & \textbf{Intensity 1} & \textbf{Intensity 2} & \textbf{Intensity 3} \\
    \midrule
    ASV19 eval & \textbf{1.71\%} & 2.72\% & 9.01\% & 7.45\% \\
    DF21 eval & 2.65\% & \textbf{2.19\%} & 3.15\% & 5.74\% \\
    DF21 hidden & 9.04\% & \textbf{6.92\%} & 11.82\% & 13.64\% \\
    LA21 eval & \textbf{8.63\%} & 10.02\% & 13.96\% & 16.23\% \\
    LA21 hidden & 12.55\% & \textbf{12.35\%} & 15.05\% & 15.97\% \\
    WaveFake & 6.46\% & \textbf{1.23\%} & 14.10\% & 5.63\% \\
    In The Wild & 3.34\% & \textbf{2.70\%} & 6.02\% & 6.85\% \\
    \bottomrule
    \end{tabular}%
    }
    \caption{Effects of data transformation intensity (EER\%)}
    \label{tab:ablation_augmentation}
\end{table}

Since transformations in the self-conversion branch must preserve speaker identity while providing sufficient acoustic variation to drive the VC process, we investigate the optimal intensity level.  As shown in Tab.\ref{tab:ablation_augmentation}, Intensity 1 achieves the best generalization, with EERs of $1.23\%$ on WaveFake and $2.70\%$ on In-the-Wild. Insufficient transformation (Intensity 0) yields higher EER on WaveFake ($6.46\%$), indicating inadequate acoustic diversity for robust artifact learning. Conversely, excessive transformation (Intensity 2–3) causes performance degradation, as the model learns transformation-specific distortions rather than universal forgery artifacts.

This analysis confirms that moderate transformation intensity optimally balances acoustic diversity with artifact preservation, which is essential for bias mitigation and cross-domain generalization.

\subsubsection{Sensitivity Analysis of Learnable Reweighting Parameters}

\begin{table}[ht]
    \centering
    \resizebox{\columnwidth}{!}{
    \begin{tabular}{lccccc}
    \toprule
    \textbf{LR} & \textbf{Alpha value} & \textbf{EER} & \textbf{ACC} & \textbf{AUC} & \textbf{AP} \\
    \midrule
    $10^{-4}$ & $w_{f}=1.73, w_{r}=0.63$ & 1.63\% & 98.37\% & 99.81\% & 99.98\% \\
    $10^{-5}$ & $w_{f}=1.77, w_{r}=0.68$ & 2.97\% & 97.02\% & 99.31\% & 99.93\% \\
    $\mathbf{10^{-6}}$ & $w_{f}=1.77, w_{r}=0.69$ & \textbf{1.23\%} & \textbf{98.77\%} & \textbf{99.90\%} & \textbf{99.99\%} \\
    \bottomrule
    \end{tabular}%
    }
    \caption{Optimization stability analysis: learning rate effects on reweighting parameters. Results are on WaveFake dataset.}
    \label{tab:ablation_lr}
\end{table}

The learnable reweighting parameters require careful optimization to maintain training stability. We analyze their sensitivity to learning rate on the WaveFake dataset. As shown in Tab. \ref{tab:ablation_lr}, a learning rate of $10^{-6}$ yields optimal performance ($1.23\%$ EER), while higher rates ($10^{-4}$, $10^{-5}$) lead to degraded results despite converging to similar weight values. This indicates that rapid weight updates destabilize the balance between real and synthetic sample gradients.

These findings validate our design choice of constraining $w\_\text{fake} \in (1, 2)$ with a $1.0$ bias: a conservative learning rate combined with bounded weight ranges ensures that synthetic samples consistently receive higher emphasis while preventing model divergence.


\subsubsection{Comparison with Conversion between Different Identities}
To validate that our same-speaker constraint effectively mitigates bias, we compare AFSS against a standard cross-speaker voice conversion baseline. In cross-speaker VC, the source and target speakers differ, creating a systematic correlation between speaker identity change and the fake label.

\begin{table}[ht]
    \centering
    \resizebox{\columnwidth}{!}{%
    \begin{tabular}{lcccc}
    \toprule
    \textbf{Method} & \textbf{ASV19-eval} & \textbf{ASV21-DF} & \textbf{WaveFake} & \textbf{In-the-Wild} \\
    \midrule
    Cross-Speaker VC & 5.68 & 21.25 & 33.11 & 55.98 \\
    \textbf{AFSS (Ours)} & \textbf{2.72} & \textbf{2.19} & \textbf{1.23} & \textbf{2.70} \\
    \bottomrule
    \end{tabular}%
    }
    \captionsetup{position=bottom}
    \caption{Comparison of Equal Error Rate (EER, \%) between our proposed AFSS and a standard cross-speaker voice conversion (VC) baseline.}
    \label{tab:sgt_vs_vc}
\end{table}

As shown in Tab. \ref{tab:sgt_vs_vc}, cross-speaker VC exhibits severe degradation on out-of-domain datasets (33.11\% on WaveFake, 55.98\% on In-the-Wild), while AFSS achieves 1.23\% and 2.70\%, respectively. This performance gap confirms that cross-speaker VC introduces speaker-related bias that fails to generalize across datasets. By enforcing identical speaker identity between real and pseudo-fake samples, AFSS eliminates this confounding factor, compelling the detector to learn universal generation artifacts rather than spurious speaker-dependent correlations.

\FloatBarrier
\section{Conclusion}
In this paper, we presented the Artifact-Focused Self-Synthesis (AFSS) method to address the bias problem and enhance the generalization capability of audio DF detection systems. By enforcing same-speaker constraints through self-conversion and self-reconstruction mechanisms, AFSS effectively eliminates confounding factors related to speaker identity and semantic content, compelling the detector to focus exclusively on authentic forgery artifacts. Integrated with a learnable reweighting loss and balanced batch sampling, our approach achieved state-of-the-art cross-domain performance with an average EER of 5.45\% across seven evaluation datasets. The high reliability of the proposed framework is evidenced by significant error reductions on WaveFake (1.23\% EER) and In-the-Wild (2.70\% EER). Most importantly, AFSS removes the dependency on pre-collected fake datasets, providing a practical and scalable solution for addressing the threats posed by rapidly evolving speech synthesis technologies.

\FloatBarrier
\newpage
\bibliographystyle{IEEEbib}
\bibliography{refs}


\end{document}